\documentclass[
]{ceurart}

\usepackage{listings}
\usepackage{array}       
\usepackage{textcomp}    

\begin{document}

\copyrightyear{2025}
\copyrightclause{Copyright for this paper by its authors.
  Use permitted under Creative Commons License Attribution 4.0
  International (CC BY 4.0).}

\conference{CHI'25 Workshop: The Third Workshop on Building an Inclusive and Accessible Metaverse for All,
   April 26, 2025, Yokohama, Japan}

\title{Self++: Merging Human and AI for Co-Determined XR Living in the Metaverse}


\author[1]{Thammathip Piumsomboon}[
orcid=0000-0001-5870-9221,
email=tham.piumsomboon@canterbury.ac.nz,
url=https://www.quadriclab.org/,
]
\cormark[1]
\address[1]{School of Product Design, University of Canterbury, Christchurch, New Zealand}

\cortext[1]{Corresponding author.}

\begin{abstract}
This position paper introduces Self++, a novel nine-level framework for co-determined living in the Metaverse, grounded in Self-Determination Theory. Self++ prioritises human flourishing by progressively cultivating competence, autonomy, and relatedness through dynamic human-AI collaboration in extended reality (XR). Unlike technologically deterministic approaches, Self++ emphasises user empowerment by enhancing competency, mitigating cognitive biases and leveraging XR’s immersive capabilities. Key research directions proposed include exploring the boundaries of user-defined AI autonomy, designing for meaningful social connection in XR, and establishing proactive ethical safeguards. Ultimately, Self++ offers a roadmap for creating a human-centred, AI-enhanced Metaverse where technology amplifies, rather than diminishes, human potential.
\end{abstract}


\maketitle

\section{Introduction} 

Picture a world where extended reality (XR) devices, such as augmented reality (AR) glasses, dissolve the barriers of distance and limitation. These devices open a collaborative frontier that transcends borders and redefines possibility. By collecting personalised data, they help weave a Metaverse where every glance and decision contributes to a universe tailored to our aspirations. Through this portal to a dynamic hybrid reality, experts, creators, and dreamers converge in a shared, immersive space. Here, every interaction and decision is amplified by artificial intelligence (AI), which synthesises vast datasets into actionable insights \cite{ball2022metaverse}. This is not just a technological leap but a reimagining of human potential, where AI evolves from tool to partner, unlocking an abundant future of unprecedented freedom. Imagine a doctor collaborating globally to transform healthcare, a child in a remote village exploring the cosmos, and an architect sculpting ideas into reality with a simple gesture. Each is empowered by a seamless blend of human ingenuity and machine intelligence.

Yet, this vision is not without peril. Following Parasuraman’s automation model \cite{parasuraman2000model}, overreliance on AI could erode the very skills that define expertise, for example, a surgeon’s tactile intuition might fade beneath algorithmic guidance, while our critical thinking could gradually be reduced to autopilot \cite{lee2025impact}. At the same time, cognitive biases such as the Dunning-Kruger effect may distort self-assessment, convincing novices they have mastered complexities they barely understand \cite{kruger1999unskilled}. Moreover, the devices enabling this fusion, such as AR glasses, threaten privacy by commodifying biometric data like heartbeats and gaze patterns, while opaque algorithms nudge users toward choices that compromise autonomy. These challenges are not hypothetical; they represent fractures in the foundation of a future that is already unfolding.

To navigate this duality, we turn to Self-Determination Theory (SDT), which posits that human flourishing hinges on autonomy, competence, and relatedness \cite{deci2012self}. The Metaverse demands frameworks that prioritise these needs, ensuring AI and XR elevate rather than diminish human agency. Autonomy as a compass guides users to steer rather than just participate, competence as craft shapes authentic mastery through AI-driven cognitive bias mitigation, and relatedness as glue strengthens connections via AI mediators that deepen collaboration across virtual and physical worlds. This is the impetus behind Self++, a 9-level framework for co-determined XR living. Grounded in SDT, Self++ reimagines design principles for the Metaverse:

\begin{itemize}
     \item Core Principles: Embedding autonomy, competence, and relatedness into system design.
     \item Competency Enhancement: Employing debiasing strategies, such as Cognitive Bias Modification, to mitigate cognitive distortions and build genuine skills.
     \item AI Partnership: Configuring AI as a dynamic, personalised collaborator that supports user growth.
\end{itemize}

This framework provides a theoretical guideline for designers and researchers to develop XR systems that enhance human potential while upholding ethical standards, drawing on preliminary research on AI-driven XR visualisation for decision support \cite{dong2024}. By operationalising these principles, Self++ aims to ensure that the Metaverse fosters equitable and empowering human experiences rather than reinforcing technological dominance.

\section{Self++: A 9-Level Framework for Co-Determined XR Living}

The Self++ framework is a 9-level model that serves as a guideline for designing human-centred AI within XR, particularly emphasising applications in the Metaverse (see Table \ref{tab:selfplusplus}). Rooted in SDT, this framework prioritises the progressive development of three fundamental psychological needs: \textbf{Competence}, \textbf{Autonomy}, and \textbf{Relatedness}. These needs are not treated as isolated elements but as interconnected drivers of user empowerment, enabling meaningful and synergistic interactions between humans and AI. Each of these core needs is subdivided into three progressive sub-levels, resulting in a nine-tier structure that spans foundational skill-building to purpose-driven engagement. Unlike traditional linear models, Self++ is conceived as a flexible and adaptive design space, capable of adjusting dynamically to the user's evolving expertise, psychological requirements, and environmental context within XR. Furthermore, it integrates insights from dual-process theory \cite{evans2013dual}, facilitating a shift from deliberate, analytical \textit{System 2} thinking in the early stages to a balanced interplay with intuitive \textit{System 1} thinking as users advance, thereby supporting both learning and fluent interaction.


\begin{table}[h]
  \centering
  \scriptsize
  \caption{Self++ Framework: 9 Levels of Co-determined XR Living}
  \label{tab:selfplusplus}
  \begin{tabular*}{\textwidth}{@{\extracolsep{\fill}}|
      >{\raggedright\arraybackslash}m{0.15\textwidth}|
      >{\centering\arraybackslash}m{0.1\textwidth}|
      >{\centering\arraybackslash}m{0.15\textwidth}|
      >{\raggedright\arraybackslash}m{0.5\textwidth}|
      }
    \hline
    \multicolumn{1}{|c|}{\textbf{Level}} & 
    \multicolumn{1}{c|}{\textbf{AI Role}} & 
    \multicolumn{1}{c|}{\textbf{Purpose}} & 
    \multicolumn{1}{c|}{\textbf{Description}} \\
    \hline
    \multicolumn{4}{|c|}{\textbf{Stage 1: Instruct - Competence Building}} \\
    \hline
    1. Guided Familiarisation & Basic Tutor & Initial Sense-Making & \textbf{Basic Competence} \textemdash Builds XR basics, eases onboarding by reducing anxiety and cognitive load, assuming limited competency to focus on skill acquisition. \\
    \hline
    2. Skill-Focused Practice & Skill-Builder & Structured Think \& Act & \textbf{Intermediate Competence} \textemdash Masters complex skills and strategies via advanced training, encouraging self-efficacy and proficiency. Competency grows with self-directed optimisation and experimentation. \\
    \hline
    3. Competency Mastery \& Refinement & Performance Coach & Advanced Think \& Act & \textbf{Advanced Competence} \textemdash Refines mastery with high competency expected, guided by AI mentorship and XR real-world simulations. \\
    \hline
    \multicolumn{4}{|c|}{\textbf{Stage 2: Empower - Autonomy \& Agency Empowerment}} \\
    \hline
    4. Choice-Supportive Nudging & Choice Architect & Gentle Guidance & \textbf{Basic Autonomy} Enhances choice with subtle guidance, offering limited autonomy via optimal strategies. \\
    \hline
    5. Collaborative Recommendation \& Justification & Informed Advisor & Shared Decision-Making & \textbf{Shared Autonomy} \textemdash Fosters shared decisions, improving autonomy with AI expertise in a collaborative setting. \\
    \hline
    6. User-Defined Autonomy \& Delegation & Customisable Agent & Flexible Control & \textbf{Advanced Autonomy} \textemdash Maximises control and human-AI synergy, enhancing autonomy through strategic AI management and workflow optimisation. \\
    \hline
    \multicolumn{4}{|c|}{\textbf{Stage 3: Connect - Relatedness \& Purpose Integration}} \\
    \hline
    7. Contextual Awareness \& Consequence Visualisation & Contextual Interpreter & Meaningful Action & \textbf{Basic Relatedness} \textemdash Links actions to broader contexts for purpose, strengthening Competence with systems-level thinking. \\
    \hline
    8. Collaborative \& Socially Aware & Social Facilitator & Collaborative Action & \textbf{Enhanced Relatedness} \textemdash Builds teamwork and connection, enhancing Autonomy and Competence in human-AI collaboration. \\
    \hline
    9. Purpose-Driven \& Meaningful Engagement & Purpose Amplifier & Meaningful Action \& Reflection & \textbf{Deep Relatedness} \textemdash Drives purpose and motivation via value alignment, fostering impactful XR living. \\
    \hline
  \end{tabular*}
\end{table}


\subsection{Stage 1---Instruct: Competence Building (Levels 1-3): Establishing a Robust Foundation}

\textbf{Competence} Building, the initial stage, lays the groundwork for user empowerment in XR. Leveraging AI and XR, this stage delivers immersive, supportive learning focused on foundational skills: clear perception, procedural fluency, and environmental comprehension. Users are guided to consciously acquire and refine these abilities through deliberate \textit{System 2} thinking, establishing a robust human information processing (sense-think-act) for future autonomy and agency in the Metaverse.

\begin{itemize}
    \item \textbf{Level 1: Guided Familiarisation (AI as Basic Tutor --- Initial Sense-Making)} \\
    \textit{Basic Competence --- Foundational Understanding and Confidence:}
    For any new journey in the Metaverse, AI can serve as a Basic Tutor by providing structured, step-by-step guidance in XR to ensure smooth onboarding. XR’s immersive capabilities enable guided virtual tours with clear visual hierarchies and interactive elements that reduce cognitive overload for novice users. For instance, an AI-led walkthrough can introduce navigation, communication tools, and basic interactions with immediate feedback, thereby building foundational competence and user confidence by prioritising simplicity and clarity. The Virtual Triplets study \cite{zhang2024virtual} further demonstrates how AI-driven guidance in immersive virtual environments can offer structured, real-time instruction to novice users, exactly the kind of support envisioned at this level.

    \textbf{Level 2: Skill-Focused Practice (Skill-Builder - Structured Think \& Act)} \\
    \textit{Intermediate Competence - Skill Development \& Procedural Mastery.}
    As users gain familiarity, AI transitions to a Skill-Builder, providing targeted practice via XR simulations dynamically tailored to individual progress. These immersive simulations, enhanced by haptic feedback and adaptive difficulty, hone specific skills and build procedural fluency. For instance, in a virtual surgery simulation, AI adjusts the procedure's complexity based on the user's suturing and incision performance while real-time analytics track improvement. Through personalised training modules and performance-driven feedback, this level fosters user competence and a tangible sense of achievement. This approach mirrors Vygotsky’s concept of scaffolding---providing temporary, tailored support within the learner's Zone of Proximal Development (ZPD) until they can master complex tasks independently \cite{vygotsky1978mind}. In the XR context, systems like XR-LIVE exemplify this strategy by employing spatial-temporal assistive toolsets, such as auto-pause and step-by-step checklists, that break complex lab procedures into manageable segments, effectively scaffolding learners and perfectly aligning with this Level \cite{thanyadit2022xr}. As users' proficiency grows, the AI gradually withdraws its guidance, much like a construction scaffold is removed once the building is structurally sound.

    \item \textbf{Level 3: Competency Mastery \& Refinement (Performance Coach - Advanced Think \& Act)} \\
    \textit{Advanced Competence - Expertise, Efficiency, \& Strategic Insight.}
    At this advanced stage, AI evolves into a Performance Coach that offers expert feedback and detailed analytics within dynamic scenarios. Users face complex, high-stakes challenges, such as managing a virtual crisis response team, while the AI delivers strategic insights, analyses decision-making patterns, suggests optimisations, and provides debiasing prompts to refine judgment. This level emphasises metacognition, empowering users to master skills and develop effective strategies in the virtual domain, ultimately achieving true competency. Moreover, by integrating diverse debiasing interventions, such as those outlined by Morewedge et al. \cite{morewedge2015debiasing} or Cognitive Bias Modification techniques \cite{hertel2011cognitive}, the system actively cultivates deliberate, analytical (System 2) thinking, even when high competency might otherwise encourage reliance on fast, intuitive (System 1) processes, thereby mitigating cognitive biases and promoting metacognitive awareness.

\end{itemize}


\subsection{Stage 2---Empower: Autonomy \& Agency Empowerment (Levels 4--6): Cultivating Self-Direction and Control}

With the solid foundation of user competence established in the initial levels, Stage 2 shifts focus to empowering user \textbf{Autonomy} and agency across Levels 4 through 6. This critical progression recognises that a sense of self-direction and control is essential for fostering intrinsic motivation and sustained engagement within the Metaverse. These levels correspond to the core ``Think'' and ``Act'' stages, encouraging users to balance intuitive \textit{System 1} thinking with deliberate \textit{System 2} control for complex decisions requiring nuanced human judgement.

\begin{itemize}
    \item \textbf{Level 4: Choice-Supportive Nudging (Choice Architect --- Gentle Guidance)} \\
    \textit{Basic Autonomy --- Guided Choice \& Subtle Influence.} 
    At this pivotal level, AI acts as a subtle Choice Architect, guiding user decisions through contextually relevant interventions while ensuring that the user remains the ultimate decision-maker. XR technology excels at delivering visually subtle \& context-aware nudges \cite{Thaler2008}, seamlessly integrating gentle AR overlays and visual cues that grasp user's attention toward beneficial options without feeling directive. The AI leverages spatial choice architectures to present options within the user's environment, highlighting advantageous choices through nuanced emphasis and intuitively visualising complex trade-offs. Importantly, the system maintains transparency by clearly indicating when it is providing nudges or suggestions, thereby safeguarding user autonomy and fostering trust. For example, the Ex-Cit XR study \cite{piumsomboon2022ex} proposed nudging by subtly modifying the immersive virtual environment, such as introducing minor visual adjustments, to gently steer users toward compliance without detracting from their overall experience. This approach aligns with the concept of nudging, wherein small, unobtrusive adjustments guide behaviour without restricting choices. 
    
    \item \textbf{Level 5: Collaborative Recommendation \& Justification (Informed Advisor --- Shared Decision-Making)} \\
    \textit{Shared Autonomy --- Collaborative Decision-Making \& Informed Choice.}
    At this level, AI evolves into an Informed Advisor that provides proactive, transparent recommendations and clear justifications, fostering a true partnership with the user. XR technology creates shared workspaces that enable direct human-AI collaboration in a blended reality, where virtual agents can interact with users as if they were colleagues. Here, AI leverages spatial visualisation to overlay data, confidence levels, and logical rationale directly onto the task context, allowing users to explore alternatives and make informed decisions collaboratively. For example, Han \cite{Han2024triad} introduces a system that integrates a large language model (LLM)-driven embodied virtual agent into a multiuser virtual environment---capable of offering recommendations, explains its reasoning, and facilitates dialogue through multimodal channels (both verbal queries and non-verbal interactions)---thereby enhancing co-presence and engagement among participants. As demonstrated by Zhang et al. \cite{zhang2021ideal}, people expect AI to exhibit human-like communication, shared understanding, and instrumental skills.

    \item \textbf{Level 6: User-Defined Autonomy \& Delegation (Customisable Agent --- Flexible Control)} \\
    \textit{Advanced Autonomy --- User-Defined Control \& Flexible Delegation.}
    At this pinnacle of autonomy empowerment, AI becomes a truly Customisable Agent, granting users unprecedented granular control over AI autonomy and enabling dynamic task delegation. XR provides intuitive interfaces that
    allow users to adjust AI assistance levels across diverse tasks. Users can define personalised automation rules and tailor AI behaviour through visual and spatial programming, while real-time dashboards enable monitoring of AI activities and seamless switching between AI-assisted and manual operation. For example, Acharya et al. \cite{acharya2025agentic} introduce their Agentic AI, which autonomously pursues complex objectives with minimal human intervention, and Yang et al. \cite{yang2025magmafoundationmodelmultimodal} demonstrate how a multimodal foundation model like Magma processes diverse inputs to execute actions in both digital and physical environments, thereby exemplifying the flexible and granular control envisioned in this level.

\end{itemize}


\subsection{Stage 3---Connect: Relatedness \& Purpose Integration (Levels 7--9): Fostering Meaningful Self++ Living}

Building upon established competence and autonomy, Stage 3 focuses on \textbf{Relatedness} across Levels 7 through 9. This critical progression connects users to broader contexts, collaborative communities, and meaningful purpose, transforming individual capabilities into deeply fulfilling experiences within the Metaverse and beyond. These levels foster connection to others, to values, and to meaningful impact, culminating in truly co-determined XR living.

\begin{itemize}
    \item \textbf{Level 7: Contextual Awareness \& Consequence Visualisation (Contextual Interpreter --- Meaningful Action)} \\
    \textit{Basic Relatedness --- Contextual Understanding \& Consequence Awareness.}

   At this stage, AI evolves into a sophisticated Contextual Interpreter that enhances user understanding of the broader implications of their actions. XR leverages overlays to contextualise virtual activities within real-world environments and societal frameworks. AI provides 3D visualisations of system impacts and interdependencies, illustrating how user decisions create ripple effects across interconnected systems. Through immersive scenario exploration, users can experience the long-term ramifications of their choices in emotionally resonant simulations, fostering a deeper appreciation of consequences beyond immediate outcomes. For example, Jain et al. \cite{jain2025visualizing} demonstrate that presenting hierarchical causality levels—such as event-level, interaction-level, and gesture-level—within XR enhances user comprehension and task execution in complex assembly tasks, aligning with this level's focus on contextual awareness and consequence visualisation.
    
    \item \textbf{Level 8: Collaborative \& Socially Aware (Social Facilitator --- Collaborative Action)} \\
    \textit{Enhanced Relatedness --- Facilitating Collaborative Action \& Strengthening Social Connection.}
    At this level, AI serves as a Social Facilitator, promoting dynamic collaboration within shared Metaverse environments. XR creates persistent virtual spaces designed specifically for sustained collaboration, enabling users to build lasting relationships and fostering a robust sense of community both in virtual interactions and through subsequent real-world connections. The system uses realistic avatars and enhanced social presence, achieved via spatially accurate audio, subtle non-verbal cues, and shared virtual body language, to make collaborative experiences feel authentically human and emotionally engaging. Furthermore, AI delivers detailed visual feedback (e.g., dashboards or immersive visualisations) of team progress and community impact, highlighting individual contributions to collective successes, thus reinforcing shared purpose and mutual accountability. For instance, Han \cite{Han2024triad} introduced an embodied virtual agent serving as a mediator, improving collaborative behaviours and teamwork efficiency. Other studies \cite{piumsomboon2018mini,piumsomboon_shoulder_2019,hart2021manipulating} have demonstrated that XR systems can adjust or customise user avatars to enhance communication effectiveness and strengthen emotional connections.
    
    \item \textbf{Level 9: Purpose-Driven \& Meaningful Engagement (Purpose Amplifier --- Meaningful Action \& Reflection)} \\
    \textit{Deep Relatedness --- Purpose-Driven Engagement \& Meaningful Impact.}
    At the apex of the spectrum, AI becomes a Purpose Amplifier, connecting user actions to meaningful outcomes and values. XR supports interactive narratives and storytelling for purpose articulation, weaving compelling stories that link even routine tasks to inspiring frameworks aligned with human values. AI offers personalised purpose \& value alignment feedback by connecting actions to personal values and broader societal goals, thereby reinforcing intrinsic motivation. The system also facilitates reflective tools and meaning-making spaces, such as structured self-reflection modules and visual mind-mapping, while visualising positive social impacts to demonstrate how virtual or physical contributions affect real-world outcomes for individuals, communities, or global challenges. For example, Yousefi et al. \cite{yousefi2024advancing} surveyed past research on the impact of embodied virtual agents within XR environments on promoting prosocial behaviours, such as empathy, cooperation, and altruism, by linking user actions to meaningful outcomes and values. By fostering prosocial behaviours, these agents could help users engage in purpose-driven activities that resonate with their personal values and contribute positively to society, which is central to this level’s emphasis on deep relatedness and meaningful impact.

\end{itemize}


\section{Discussion and Future Work---Navigating the Co-Determined Metaverse}

The Self++ framework moves beyond simply \textit{using} AI in the Metaverse; it proposes a \textit{co-determined} existence, where human agency and AI capabilities are interwoven. This isn't about automation, but about a dynamic partnership, guided by SDT's core tenets: Competence, Autonomy, and Relatedness. The 9-level structure isn't a rigid ladder, but a flexible ``design space'' adapting to the user's evolving needs and the context of their XR experience in the Metaverse.

Key considerations of Self++, along with crucial research directions, are outlined below:

\begin{itemize}
    \item \textbf{Human-Centred Progress and Dynamic AI Agency:} Self++ resists technological determinism, prioritising human flourishing (defined by SDT) over mere technical advancement.  It also avoids simplistic views of AI agency, proposing a nuanced spectrum of human-AI collaboration. \textit{Future work must determine how to measure Metaverse ``progress'' by human well-being, not just technical capability, and investigate the ideal ``hand-off'' points between human and AI control, ensuring seamless transitions in XR interfaces.}

    \item \textbf{Cognitive Bias Mitigation and XR Empowerment:} Self++ proposes integrating debiasing strategies to counter the cognitive biases and foster accurate self-awareness, while leveraging XR's unique capabilities (immersion, spatial interaction) to enhance user growth. \textit{Research should explore how to design XR environments that not only mitigate bias but also cultivate metacognitive skills, and how to move beyond replicating real-world tasks to create novel forms of human-AI collaboration using XR's affordances.}

    \item \textbf{The Ethical Imperative and Proactive Frameworks:} While Self++ offers a strong ethical grounding, the potential for unintended consequences in the Metaverse remains, as summarised by Slater et al. \cite{10.3389/frvir.2020.00001}. Data privacy, algorithmic bias, and manipulation require constant vigilance, alongside proactive ethical frameworks. \textit{We must determine how to build and maintain ``ethical guardrails'' in the Metaverse, ensuring Self++ principles are actively implemented and enforced.}

     \item \textbf{Empirical Validation, Autonomy, and Relatedness:} Rigorous testing is crucial to validate the 9-level progression and measure its long-term impact.  Level 6 (User-Defined Autonomy) and Levels 7-9 (Relatedness) represent significant research frontiers. \textit{Future work must identify meaningful metrics for evaluating co-determined XR systems (measuring ``flourishing''), explore how users interact with customisation AI agents, balance user control with AI support, and design for genuine empathy, collaboration, and shared purpose in XR.}

    \item \textbf{Longitudinal Studies:} Finally, the long-term effects of continued XR and Metaverse use on individuals are not well understood. \textit{Research must investigate how increasing XR/Metaverse use affects an individual's autonomy, competence, and relatedness in the real world.}
\end{itemize}

By embracing these challenges, we can move towards a Metaverse that is not just technologically advanced but truly human-centred, a co-determined space where technology and human potential amplify each other.

\bibliography{main}

\end{document}